A Review of Methods and Practices for Missing Data in Sequential Multiple Assignment Randomized Trials (SMARTs): An Ancillary Study of a Scoping Review

Nikki L. B. Freeman[1,2,3], Chenyao Yu[4], Margaret Hoch[3,5], Sydney E. Browder[6], Bradley G. Hammill[2,7,8], Avi Kenny[1,9], Kevin J. Anstrom[5,10], and Michael R. Kosorok[3,5]

1. Department of Biostatistics and Bioinformatics, Duke University, Durham, NC
2. Duke Clinical Research Institute, Duke University, Durham, NC
3. Center for A.I. and Public Health, University of North Carolina at Chapel Hill, NC
4. Department of Statistical Science, Duke University, Durham, NC
5. Department of Biostatistics, University of North Carolina at Chapel Hill, Chapel Hill, NC
6. Department of Epidemiology, University of North Carolina at Chapel Hill, Chapel Hill, NC
7. Department of Population Health Science, Duke University, Durham, NC
8. Department of Medicine, Duke University, Durham, NC
9. Duke Global Health Institute, Duke University, Durham, NC
10. Collaborative Studies Coordinating Center, University of North Carolina at Chapel Hill, NC

Corresponding author
Nikki L. B. Freeman
Duke Clinical Research Institute
300 W. Morgan
Durham, North Carolina 27701
nikki.freeman@duke.edu

Word count: 4730
Table count:4
Figure count: 11
References count:
Supplement: 2 documents, 5 tables

# Abstract

**Background:** Missing data poses an acute threat to sequential multiple assignment randomized trial (SMART) analyses because of the sequential treatment structure and response-dependent re-randomization.

**Objectives:** This study aimed to (1) review the current statistical methods for handling missing data in SMARTs, and (2) characterize how missing data is reported and handled in published SMARTs.

**Methods:** We conducted a narrative review of statistical methods developed for missing data in SMARTs. Additionally, we conducted a pre-specified secondary extraction of a previously published scoping review of SMARTs focused on missing data. Extraction captured attrition rates, methods for handling missingness, and planned versus performed missing data analyses.

**Results:** Seven methodological papers were identified; nearly all assume missing at random (MAR), and only one addresses the full set of SMART-specific missingness types. Across 30 published SMARTs, median overall attrition was 18.1% (range 0.6%–56.5%). Methods used to address missing data were described in 80% of the manuscripts; mixed-model methods were most common (30%). Among 14 studies with paired protocols, sensitivity analyses were pre-specified in 2 (14%).

**Conclusions:** SMART-specific methodology for missing data is limited, and a substantial gap exists between available methodology and current SMART practice.

**Introduction**

Sequential multiple assignment randomized trials (SMARTs) are clinical trial designs in which participants may be randomized to treatments at two or more sequential decision points, with later randomizations potentially depending on a participant's response to earlier treatments.[1,2] SMARTs are flexible trial designs, and data from SMARTs can be used to test hypotheses about first- and second-line therapies.[3] Moreover, by formalizing the sequential treatment decision process, SMARTs generate data uniquely suited for testing and estimating dynamic treatment regimes (DTRs). DTRs are sequences of decision rules that map from patient characteristics and treatment history to treatment recommendations.[3–7] SMARTs have been employed across a wide range of clinical domains, including behavioral and mental health, infectious disease, oncology, and pain management.[8–10]

A recent scoping review of SMARTs in human health research identified 89 studies (59 protocol papers, 16 primary analysis papers, and 14 studies with both), revealing substantial growth in the use of SMART designs, but also significant variability in how these trials are reported.[8] While that review focused on design characteristics and reporting practices broadly, one critical dimension of SMART reporting, missing data, has not been systematically examined. This gap is significant because missing data poses an acute threat to SMART analyses for reasons that go beyond the standard concerns about incomplete data in clinical trials.

In a conventional two-arm randomized controlled trial, the primary concern with missing data is typically the loss of the primary outcome for some participants, which can lead to potential bias and reduced power. In SMARTs, the problem is structurally more complex. Data may be missing not only for the final outcome but also for the intermediate responder/nonresponder classifications that determine subsequent treatment pathways, for treatment assignments at later stages, and for time-varying covariates that may serve as tailoring variables in treatment decision rules. Each type of missingness has distinct implications depending on the analytic target, e.g., comparisons of first- and second-line treatments, evaluations of embedded DTRs, and estimation of optimal DTRs. The methodological literature on handling missing data in SMARTs, however, is sparse. Additionally, it is unknown whether these methods, or any others, are used in practice.

This paper has two objectives. First, we review the current state of statistical methods for handling missing data in SMARTs. Second, as an ancillary study of the scoping review by Freeman et al.,[8] we characterize how missing data is reported and handled in practice, focusing on the 30 published SMART primary analyses identified in that review. By placing the methods review alongside an empirical assessment of current practice, we aim to identify gaps between available methods and what is actually being done, and to provide a foundation for improved reporting standards and methodological development.

**Background**

*The SMART Design*

A SMART is a multi-stage randomized trial in which participants are randomized to treatments at two or more decision points. Randomization to secondary and tertiary treatments may depend on the response to first-line treatment. Pre-specified criteria are used to determine whether a participant is a "responder" or "non-responder." One common design is to re-randomize first-line therapy non-responders to second-line therapy, and for responders to remain on first-line therapy or be placed under surveillance, depending on the nature of first-line treatment. Second-line therapies may include continuing the first-line treatment, augmenting it, or switching to a different treatment. This structure can be extended to three or more stages and is intended to mimic real-world sequential clinical decision-making pathways.[1,2] An example of a possible SMART design from Freeman et al.[8] is reproduced in Figure 1.

[FIGURE 1. Example SMART schematic]

SMARTs are designed to address multiple interrelated analytic questions within a single trial. These include: (1) *comparing first-line treatments* by their main effects on the outcome; (2) *comparing second-line treatments* by their main effects on the outcome; (3) *comparing embedded treatment regimes*, the fixed treatment sequences that are built into the trial design (e.g., "start with A, and if nonresponder, switch to B else stay on A" versus "start with A, and if nonresponder, augment with C else continue to surveil"); and (4) *estimating optimal individualized DTRs*, treatment rules that map individual patient characteristics to recommended treatments at each stage to optimize expected outcomes.[3,5,6]

*Missingness in SMARTs*

While there are many ways to conceptualize missing data in SMARTs, e.g., using a causal inference/counterfactual direct acyclic graph-based lens, we will use a simple three-axis characterization: (1) the type of variable affected, i.e., responder status $R$, treatment assignment $A$, covariate/tailoring variable $X$, or outcome $Y$, (2) the pattern of missingness, i.e., monotone versus non-monotone, and (3) the mechanism, i.e., missing at random (MAR), missing not at random (MNAR). Missing completely at random (MCAR) is a limiting case typically handled by standard methods, and we do not focus on it here. Each combination of characteristics corresponds to a distinct identification and estimation problem.

**A Review of Statistical Methods for Missing Data in SMARTs**

In this section, we review statistical methods for handling missing data in SMARTs. We organized this review by two primary missing data approaches: (1) an imputation-based approach and (2) a weighting-based approach.

*Imputation-based Approach*

Shortreed et al.[11] proposed a time-ordered nested conditional multiple imputation strategy for SMARTs under the MAR assumption. This method addresses missing data across all variable types: $R$, $A$, $X$, and $Y$. It handles SMART-specific challenges, including irregularly timed $R$ and end-of-stage $X$ at outcome-dependent stage transitions, structurally missing R, and missing $R$, $A$, $X$, and $Y$ due to dropout before stage completion and re-randomization. This approach exploits nearly-monotone missingness from study attrition by using fully conditional specification (FCS) where separate imputation models are fit for each variable at each time point, with predictors restricted to current or earlier time points only, ensuring a coherent joint distribution. Bayesian mixed effects models are nested within FCS to impose smoothness on longitudinal $Y$. By sequentially imputing early-stage $X$ and $R$ first, this method then imputes later stage A, end-of-stage $X$, and $Y$, facilitating estimation of any estimand, including optimal DTR parameters, embedded regime comparisons, and value functions. This method assumes MAR conditional on observed data and does not address MNAR.

This time-oriented nested conditional imputation strategy aligns with SMART data generation, where earlier participant information determines subsequent data collection, making the temporal ordering intuitive. However, specifying numerous conditional models, one per variable per time point, creates substantial modeling burden and misspecification risk. The approach is fully parametric and, to our knowledge, lacks publicly available SMART-specific software. Additionally, the MAR assumption may be violated if missingness depends on unobserved factors.

While not a new imputation method, Shen et al.[12] proposed two post-imputation inference frameworks for single stage individual treatment rules (ITRs) under the MAR assumption when $X$, $A$, or $Y$ are missing. The first is a data-splitting approach that divides data into training and testing sets, imputes separately to avoid data leakage, estimates an optimal ITR via model-averaged Q-learning on training imputations, and evaluates its conditional value on test imputations using inverse probability weighting (IPW) or augmented IPW (AIPW) combined via Rubin's rule for asymptotic confidence intervals. The second is an $m$-out-of-$n$ bootstrap approach, which imputes the full dataset, estimates an ITR, and constructs confidence intervals using a modified $m$-out-of-$n$ bootstrap, extending the work of Chakraborty et al.,[13] who showed that the $m$-out-of-$n$ bootstrap yields consistent inference for the non-smooth value functional of a data-driven DTRs. The modification estimates resample size $m$ per imputation using Chakraborty et al.'s data-driven formula, takes the minimum $m$, constructs intervals per imputation, and averages percentiles. Both approaches assume MAR with some robustness to weak MNAR. The $m$-out-of-$n$ bootstrap is preferred for small samples because the full data set can be used to estimate optimal ITRs, whereas data-splitting is preferred for large datasets for computational efficiency.

***Weighting-based Approach***

Dong et al.[14] proposed an AIPW approach for estimating optimal DTRs under monotone missingness, with applications including SMARTs with missing data. This method addresses missingness in $X$, $A$, and $Y$ while assuming baseline variables ($X_1$, $A_1$) are always observed. It requires a monotone missingness pattern, and sparse non-monotone missingness is handled through artificial censoring or single imputation. This approach assumes MAR, where dropout probability depends only on observed history, modeled via discrete hazard functions using logistic regression. The target estimand is the optimal regime maximizing mean outcome $V(\pi)$. The key advantage is double robustness, where the estimator remains consistent if either the dropout hazard model or the conditional expectation model is correctly specified.

The AIPW method offers notable strengths, including double robustness, requiring only one of two models to be correct for consistency. However, the method has an untestable MAR assumption that fails under MNAR scenarios. Additionally, high-dimensional trajectory spaces can produce unstable solutions requiring regularization techniques that introduce tuning complexity and potential bias-variance tradeoffs.

Huang and Zhou[15] proposed an AIPW approach to estimate optimal treatment-selection rules when $X$ is partially missing, while $Y$ and $A$ are fully observed. This method does not specifically address SMART trials; rather, it focuses on single-stage treatment decisions in cohort studies or randomized trials. The approach handles general missingness in covariates under the MAR assumption. The target estimand is the optimal treatment-selection rule that minimizes the population disease rate. The proposed AIPW estimator achieves double robustness by modeling four components: (1) missing data mechanism, (2) calibration model, (3) propensity score, and (4) outcome regression. The estimator remains consistent if either the missing data or calibration model is correct and either the propensity score or outcome regression model is correct.

This AIPW method offers notable strengths including double robustness requiring correct specification of only one model from each pair (missing data or calibration and propensity score or outcome regression). Simulation studies demonstrate consistent superiority over complete-case and simple IPW approaches, particularly with substantial missingness, and the method flexible handles discrete and continuous covariates and multiple covariates missing together within a unified direct-optimization framework. However, key limitations include the requirement to specify four working models, which may lead to incompatibility issues if the calibration models conflict with directly specified models, though flexible semiparametric approaches can mitigate this. The MAR assumption is strong, and the method is restricted to single-stage treatment decisions. The method assumes treatment and outcome are fully observed, and solving weighted classification problems may face computational challenges in high-dimensional settings.

Sun et al.[16,17] wrote two papers that extend optimal DTR estimation to observational cohort data with MNAR covariates, representing a notable departure from the MAR assumption in the SMART-specific methods we reviewed. Their approach is motivated by MNAR arising from informative patient monitoring where missingness depends on unobserved covariate values, creating a unique DTR challenge: backward-induction-induced missing pseudo-outcomes where even fully observed outcomes $Y$ lead to MNAR pseudo-outcomes at earlier stages because they depend on MNAR covariates from later stages. Under a future-independent missingness assumption, they developed weighted Q-learning using semiparametric missingness models identified via nonresponse instrumental variables, and proposed more robust Covariate-

Functional-Balancing Learning (CFBL) and Augmented CFBL (ACFBL) methods that use RKHS-based balancing weights and nonparametric Q-functions to mitigate misspecification risk. While not SMART-specific, these methods address covariates MNAR in optimal DTR estimation.

*Summary of Methods*

Six prior papers discussed statistical methods for missing data in SMARTs or related to optimal DTR estimation. **Table 1** summarizes the setting, mechanism, pattern, approach, and missing data elements for each paper.

**Missing Data in Published SMARTs**

*Methods*

The parent scoping review for this study identified SMARTs for human health published between 1 January 2009 and 9 February 2024[11]. Pubmed (U.S. National Library of Medicine, National Institutes of Health), Scopus (Elsevier), Embase (Elsevier), and the Cochrane Central Register of Controlled Trials (CENTRAL) (Cochrane Library) were searched using a combination of controlled vocabulary, where applicable, and various keywords for SMARTs. The protocol for this review was consistent with the PRISM-ScR guidelines and was posted publicly before the beginning of the review[8]. In this review, 5486 studies were screened and 103 met the inclusion criteria.

For this ancillary analysis, only the 30 studies with a primary analysis paper were included; 14 had a corresponding protocol paper, allowing comparison between planned and executed methods for handling missing data. In the parent scoping review, two fields related to missing data were included for the protocol-only SMARTs; no secondary extraction was completed for those. Results of those two fields have not previously been reported and are included in this manuscript.

The secondary extraction followed an analysis plan pre-registered on OSF.[18] Domains included in the expanded extraction form included attrition and study flow, types of missing data reported, missing data mechanism, methods used to handle missing data, sensitivity analyses, reporting and discussion of missingness, and, for those studies with both a protocol and a primary analysis paper, concordance between the protocol and the primary analysis paper. The pre-specified analysis plan and full extraction form are available in the supplemental materials (**Supplementary documents 1 and 2**).

Relevant fields in the extraction form were pre-populated based on the parent scoping review. For initial data extraction, the large language model Claude Opus 4.6 extended thinking (Anthropic 2026) was used. A standardized prompt was used, which specified all extraction fields, their definitions, response options, and instructions for handling ambiguous cases; the model returned structured output in JSON format. Before full-scale extraction, we calibrated the AI-assisted approach on a subset of six papers. For these, an author independently performed full manual extraction and compared the two extractions field by field. Based on these comparisons, fields were assigned reliability tiers: high for concordance in 5 or more papers, moderate for concordance in 4 papers, and low for concordance in 3 or fewer papers. For full extraction, moderate- and low-reliability fields were verified or manually extracted by an author for every paper, and high-reliability fields were checked in a random 20% sample of the papers. Additionally, a second author verified key fields in a random sample of six papers: the primary missing data method, whether a SMART-specific method was used, whether sensitivity analyses were conducted, and the overall attrition rate. Protocol concordance was assessed by a human reviewer without the use of AI.

We present appropriate descriptive statistics for each extraction field. For categorical fields, these include frequencies and percentages; for continuous fields, these include median and range.

**Results**

*Study Characteristics*

This ancillary analysis included 30 published SMART primary analysis papers. Of these, 16 were primary analysis papers without a corresponding protocol paper, 14 had both a protocol and a primary analysis report, and 10 were pilot studies. The studies covered a broad range of therapeutic areas. Mental and behavioral health was the most common area, with 11 studies, followed by substance use, with 8 studies. The remaining studies were distributed across overweight and obesity, cancer, HIV, sleep, pain, osteoarthritis, and neurological disease or injury. Nearly all studies (29 of 30) used a two-stage SMART design, while one used a three-stage design. The number randomized at stage 1 varied widely across studies, ranging from 18 to 1,809. High-study characteristics are described in **Supplementary Table 1**; full details of the analytic sample are described in the parent scoping review.[8]

*Attrition and study flow*

Attrition reporting is summarized in **Table 2**. Most reviewed studies reported a CONSORT flow diagram (28/30, 93.3%), with similar proportions among pilot (9/10) and full-scale studies (19/20). Reporting of overall study attrition was less consistent. Twenty-four studies (80%) reported a rate that could be unambiguously extracted, while 6 (20%) were indeterminant because the language used in the text or CONSORT diagram did not clearly distinguish between treatment adherence, missing outcome measurement, and dropout from the study. For example, some CONSORT diagrams reported "completers" and "non-completers" without specifying whether these categories referred to completion of the assigned therapy or of the study. Among the 24 studies with extractable rates, the median overall attrition rate was 18.1% (range: 0.6%-56.5%), with comparable rates across pilot (median 18.6%) and full (median 17.5%) studies. Wider dispersion was observed in the full-study group.

Attrition reporting between the first and second stages was generally present. Of the 28 studies for which between-stage attrition was applicable (2 studies reported first-stage results only), 25 (89.3%) reported between-stage attrition in a form that could be extracted, 2 (7.1%) did not report it, and 1 (3.6%) was indeterminant. Among studies with extractable rates, the median attrition between the first and second stages was 9.8% (range: 0%-34.8%), with similar medians across pilot (10.8%) and full (9.8%) studies. Details from the extraction are in **Supplementary Table 2.**

*Missing Data Mechanism*

Most reviewed papers (14/30) did not explicitly discuss the missingness mechanism; half of these were pilot studies. Six papers (1 pilot, 5 primary analyses) explicitly acknowledged making the MAR assumption in their analyses. For the remaining 10 papers, the missingness mechanism could be inferred from the methods used or cited. This included 7 papers that relied on MAR and 3 that used conservative strategies consistent with MNAR. Among the 13 papers that implicitly or explicitly assumed MAR, 6 provided justification for their assumption. Justification strategies ranged from brief assertions supported only by low missingness rates or balanced attrition across arms to more substantive defenses involving covariate comparisons, design-based arguments about auxiliary variables, or sensitivity analyses for MAR violations. Details of missing data mechanism reporting are in **Supplementary Table 3**.

*Methods Used to Handle Missing Data*

**Table 3** describes the methods used to handle missing data in the studies reviewed. Missing data methods were explicitly described in 24 studies (80%), with full-scale studies more likely than pilot studies to do so (85% versus 60%). Mixed models or other likelihood-based approaches were most common (9 studies; 30%), followed by multiple imputation (6 studies; 20%), and single imputation (4 studies; 13.3%). Complete case analysis was explicitly described in 2 studies (6.7%), but it is likely that it was also employed in the 7 studies (23.3%) that did not explicitly report a strategy for handling missing data. Only 1 of the 30 studies (3.3%) used a missing data method specifically developed for the SMART setting, citing Shortreed et al.[11] in its description of the multiple imputation approach used. Among the 17 full studies that described a missing data method, 10 (58.8%) reported additional analyses, e.g., sensitivity analyses, related to missing data.

*Planned Analyses and Protocol Concordance*

Among the 14 studies with both a published protocol and a primary analysis paper, plans for handling missing data were described in the protocol for 6 studies (42.9%), all of which were full-scale trials; none of the 4 pilot studies specified a plan (**Table 4**). Sensitivity analyses for missing data assumptions were pre-specified even less often, appearing in only 2 protocols (14.3%), again exclusively among full studies. Planned approaches included likelihood-based methods that can handle missing observations, examinations of missingness mechanisms, single- and multiple-imputation, and explicit plans not to use imputation.

Concordance between planned and implemented missing data approaches could be assessed for the 6 studies whose protocols addressed missing data. Only 1 study (16.7%) showed full concordance; the remaining 5 showed partial concordance. Departures from planned analyses included sensitivity analyses that were not reported in the primary analysis paper, and multiple imputation or sensitivity analyses that appeared in the primary analysis paper without having been pre-specified in the protocol. Full details are described in **Supplementary Table 5**.

**Discussion**

This study placed two views of missing data in SMARTs side by side: a narrative review of available methodological work and a pre-registered secondary extraction of how 30 published SMART primary analyses reported and handled missing data. Two findings dominate. First, the methodological literature on missing data in SMARTs (or closely related optimal DTR estimation) is sparse and almost uniformly grounded in the missing-at-random (MAR) assumption. Second, this small literature is largely not used in practice. SMART analyses overwhelmingly rely on conventional clinical trial methods, missing data mechanisms are seldom discussed, and the reporting of missingness is often insufficient even to determine its extent, especially for later SMART stages.

*The methodological landscape*

We identified only seven papers that develop or extend methods for missing data in the SMART setting or for closely related dynamic treatment regime estimation. The Shortreed et al.[11] multiple imputation strategy remains the only published method we could identify that addresses the full set of variable types that can be missing in a SMART, responder status (R), treatment assignment (A), tailoring/covariates (X), and outcomes (Y). Subsequent contributions have addressed adjacent problems: AIPW estimation for monotone missingness in optimal DTR estimation,[14] AIPW for missing tailoring covariates in single-stage rule estimation,[15] post-imputation inference for the value of an estimated rule,[12] and recent extensions to MNAR settings for observational DTR estimation with informative covariate monitoring.[16,17] The collective scope of this literature, however, is narrow. With the exception of Shortreed et al., none of the methods address missing responder status, even though R is arguably the most structurally disruptive type of missingness in a SMART because it determines downstream re-randomization and renders standard doubly robust estimators undefined. Methods for non-monotone missingness in R or A are similarly absent. MNAR-targeted methods are emerging, but the most recent contributions focus on observational rather than SMART data. Finally, we found no publicly available SMART-specific software for any of the reviewed methods.

### *Reporting practices*

The empirical picture is consistent with the methodological literature's small footprint. Across 30 SMART primary analyses, missing data were generally non-trivial; median overall attrition was 18.1%, with several studies exceeding 30% and one approaching 57%. Yet the missingness mechanism was not discussed in nearly half of the studies. When MAR was assumed, explicitly or implicitly, the most frequent justifications appealed to low overall missingness or to balanced attrition across arms. Mixed-model and likelihood-based approaches were the most common analytic strategy. While these methods can be valid under MAR for the outcome, they neither speak directly to missing *R* or *A* nor address the response-dependent stage transitions that make SMART data structurally distinct from a conventional longitudinal trial.

A specific reporting failure deserves attention. In 20% of studies, the overall attrition rate could not be unambiguously extracted because dropout from the study, nonadherence to the assigned therapy, and missing post-randomization measurement were not clearly distinguished in the text or CONSORT diagram. In a conventional two-arm trial this conflation is a nuisance; in a SMART it is a structural problem because each of these phenomena maps to a different type of missingness in the analytic data, with different identification implications. A participant who completes first-line treatment but is lost before $R$ is assessed produces missing $R$; a nonadherent participant who is followed and re-randomized produces a different counterfactual question altogether; a participant with missing $X$ at the stage transition presents a third distinct identification problem. Without disambiguation, downstream missingness cannot be coherently classified, and the appropriate analytic strategy cannot be selected.

*A methods–practice gap*

Only 1 of 30 studies (3.3%) used a missing-data method developed for the SMART setting. Several factors plausibly contribute. The available SMART-specific methods are relatively new, having been developed within the last 15 years, and lack accompanying software, which raises the cost of use even when their existence is known. SMARTs to date have concentrated in mental and behavioral health and substance use research, where mixed-model conventions for handling missing outcomes are entrenched and may be carried into SMART analyses without explicit consideration of SMART-specific missingness types. Reviewer expectations may not yet reflect the structural complexities of SMART data.

The protocol-to-paper concordance results reinforce these patterns. Among the 14 studies with both a protocol and a primary analysis paper, only 6 protocols specified a missing data plan, only 2 pre-specified a sensitivity analysis, and complete concordance was observed in just 1 study. Where protocols and analyses diverged, multiple imputation and sensitivity analyses were as often added on the basis of post-hoc judgment as they were dropped from the analysis plan, an asymmetric pattern that suggests missing data is being treated as a downstream contingency rather than as an analytic question to be specified up front.

A further factor specific to the SMART setting is sample size. Among the full (non-pilot) studies we reviewed, the median number of participants randomized at stage 1 was 115 (range [18, 1809]). Re-randomization to stage 2 further fragments the analytic sample across the response-by-treatment cells that define embedded regimes and downstream estimands. Several SMART-relevant missing-data methods are data-intensive even before this fragmentation. Multiple imputation strategies of the form proposed by Shortreed et al.[11] require fitting separate conditional models for each variable at each time point, with predictors drawn from the full observed history. Even in modestly sized SMARTs, these per-cell models can become unstable or fail to converge. Doubly robust methods[14,15] require the correct specification of one of two nuisance models, both of which must be estimable within the available sample; with thin response-by-treatment cells, propensity, hazard, and outcome regression models can have wide standard errors or break down entirely. Methods for MNAR settings[16,17] typically rely on additional structure, such as instrumental variables, balancing weights, or sensitivity parameters, which impose further demands on the data. The practical consequence is that even teams aware of SMART-specific methods may reasonably default to simpler, more numerically stable approaches in trials where the available data cannot support the full machinery. This points to a methodological gap that is not merely about availability or software, but about the need for methods that remain reliable when sample sizes are modest and analytic cells are sparse, conditions that characterize a large share of current SMART practice.

*Recommendations*

Several practical steps would narrow the gap between methods and practice without waiting for new methodological work. First, study reports and CONSORT diagrams should disambiguate dropout from treatment nonadherence and from missing measurement, ideally at each stage of the SMART. Second, missing data should be reported in terms that recognize the SMART setting. For example, we are partial to the simple taxonomy of identifying what is missing (R, A, X, Y), the pattern (monotone vs. non-monotone), and the assumed mechanism. Along with better reporting, explicit, substantive justification for the assumed mechanism should be provided. Third, missing data approaches and sensitivity analyses should be pre-specified in protocols and statistical analysis plans rather than introduced ad hoc; even simple, accessible sensitivity analyses (e.g., delta-tilting or pattern-mixture variants) materially strengthen the credibility of primary results. Fourth, SMART-appropriate methods should be considered when applicable. The default of mixed-model analyses should be paired with explicit attention to whether they appropriately handle the relevant SMART-specific missingness types, or supplemented by methods that do. These steps are consistent with calls for more complete reporting of SMARTs more generally,[8,10] and would help align analytic practice with the structural complexities of the design.

***Strengths and limitations***

This work has several strengths. To our knowledge, it is the first study to combine a focused review of methodological work on missing data in SMARTs with a structured empirical assessment of how missing data is reported and handled in published SMART analyses. The empirical extraction was conducted under a pre-registered analysis plan, with an extraction form and field definitions specified before review. The use of an AI-assisted extraction was paired with a calibration step on a six-paper subset, assignment of reliability tiers per field, manual extraction or verification of moderate- and low-reliability fields for every paper, random verification of high-reliability fields, and independent human review of protocol concordance, which together support the credibility of the extracted findings. The analytic sample was drawn from a recently completed and broadly scoped scoping review.[8]

Several limitations should also be considered. First, this is a secondary analysis of published reports; we can evaluate only what the authors chose to report. Underlying analyses may have included additional considerations of missing data that did not appear in the primary papers. Second, the sample is modest (30 primary-analysis papers), and the parent search ended in February 2024, so SMARTs published or methods developed after that date are not represented. Third, our review of methods is narrative rather than a formal systematic review; we may have missed methodological work that does not surface under SMART- or DTR-specific search terms. Fourth, despite the calibration and verification steps described above, AI-assisted extraction may underweight non-standard descriptions of missing-data handling, and our findings should be interpreted with this in mind.

*Conclusion*

The methodological literature on missing data in SMARTs is limited, almost entirely grounded in the MAR assumption. Current practice does not draw on this literature, as most published SMARTs do not report using SMART-specific missing data methodology. Closing the gap will require both new methodological work, particularly for missing responder status, non-monotone patterns, and MNAR mechanisms, and software availability to facilitate the application of these methods.

Table 1. Summary of methods specific to missing data in SMARTs or closely related

| Author (Journal, Year) | Setting | Mechanism | Pattern | Approach | Missing Data Elements |
|---|---|---|---|---|---|
| Shortreed et al. (Stat Med, 2013) | SMART with K stages; any variable may be missing | MAR | Monotone | Time-ordered nested conditional MI via Bayesian linear mixed models | R, $A_2$, X, Y |
| Dong et al. (Stat Med, 2019) | SMART with K stages; any variable may be missing | MAR | Monotone | AIPW complete case estimator for optimal DTR estimation | R, $A_2$ |
| Huang & Zhou (Stat Med, 2018)* | Single-stage; missing tailoring variables | MAR | Single-stage (N/A) | Augmented IPW for missing covariates in classification-based DTR estimation | $X_1$ |
| Shen, Hubbard & Linn (Stat Med, 2022)* | Post-MI inference for treatment rules | Per MI method | Per MI method | Sample splitting and m-out-of-n bootstrap for valid inference on DTR value after MI | Inference layer applicable to any MI approach |
| Sun, Fu & Su (Biometrics, 2025)* | Multi-stage DTR estimation from observational data; missing covariates from informative monitoring | MNAR | Non-monotone (intermittent missing covariates; induced missing pseudo-outcomes) | Weighted Q-learning with semiparametric missingness propensity model and nonresponse instrumental variables; sensitivity analysis | $X_k$ |
| Sun, Fu & Su (arXiv, 2025)* | Same as above; addresses robustness to Q-function misspecification | MNAR | Non-monotone | Direct-search DTR estimation with RKHS-based balancing weights and nonparametric Q-functions; IPW for nonignorable missingness | $X_k$ |

* Not SMART-specific, but related to missingness in the estimation of optimal dynamic treatment regimes

Table 2.

| Reporting item | Pilot studies (N = 10) | Full studies (N = 20) | All studies (N = 30) |
|---|---|---|---|
| CONSORT Diagram n (%) | 9 (90%) | 19 (95%) | 28 (93.3%) |
| Reporting of overall attrition rate n (%) | | | |
|     Reported | 9 (90%) | 15 (75%) | 24 (80%) |
|     Indeterminant* | 1 (10%) | 5 (25%) | 6 (20%) |
| Overall attrition rate among reported *Median [min, max]* | 18.6 [1.25, 32.2] | 17.5 [0.6, 56.5] | 18.1 [0.6, 56.5] |
| Reporting of attrition between first and second stages n (%) | | | |
|     Reported | 8 (80%) | 17 (85%) | 25 (83.3%) |
|     Not reported | 1 (10%) | 1 (5%) | 2 (6.7%) |
|     Indeterminant* | 0 (0%) | 1 (5%) | 1 (3.3%) |
|     Not applicable** | 1 (10%) | 1 (5%) | 2 (6.7%) |
| Attrition between first and second stages among reported *Median [min, max]* | 10.8 [0, 22.6] | 9.8 [0, 34.8] | 9.8 [0, 34.8] |

*In some cases, the overall attrition rate was indeterminant because the language used in the text or CONSORT diagram did not clearly distinguish between adherence, missing measurement, or actual dropout from the study. For example, some CONSORT diagrams reported "completers" and "non-completers" but this could refer to differing levels of adherence (completed the course of therapy) or the study. In other cases, the text or CONSORT diagram were not clear enough to make confident calculations.
** Two studies only reported the results from the first stage of the SMART

Table 3. Methods used to handle missing data

|  | Pilot studies (N=10) | Full studies (N=20) | All studies (N=30) |
|---|---|---|---|
| Was any method used to handle missing data? *n (%)* | | | |
|     Yes | 6 (60%) | 17 (85%) | 24 (80%) |
|     No | 4 (40%) | 3 (15%) | 6 (20%) |
| What method was used to handle missing data? *n (%)* | | | |
|     Complete case analysis | 1 (10%) | 1 (5%) | 2 (6.7%) |
|     Single imputation | 1 (10%) | 3 (15%) | 4 (13.3%) |
|     Multiple imputation | 1 (10%) | 5 (25%) | 6 (20.0%) |
|     Mixed models/likelihood-based | 3 (30%) | 6 (30%) | 9 (30%) |
|     Inverse probability weighting | 0 (0%) | 1 (5%) | 1 (3.3%) |
|     Doubly robust | 0 (0%) | 1 (5%) | 1 (3.3%) |
|     None | 4 (40%) | 3 (15%) | 7 (23.3%) |
| Was a SMART-specific missing data method used? | 0 (0%) | 1 (5%) | 1 (3.3%) |

Table 4. Planned analyses and protocol concordance

|  | Pilot studies (N=4) | Full studies (N=10) | All studies (N=14) |
|---|---|---|---|
| Plans for missing data in protocol | 0 (0%) | 6 (60%) | 6 (42.9%) |
| Plans for missing data sensitivity analysis in protocol | 0 (0%) | 2 (20%) | 2 (14.3%) |
| Concordance between protocol and primary analysis paper (among those with a plan in the protocol) | | | |
|     Yes | 0 (0%) | 1 (16.7%) | 1 (16.7%) |
|     Partially | 0 (0%) | 5 (83.3%) | 5 (83.3%) |

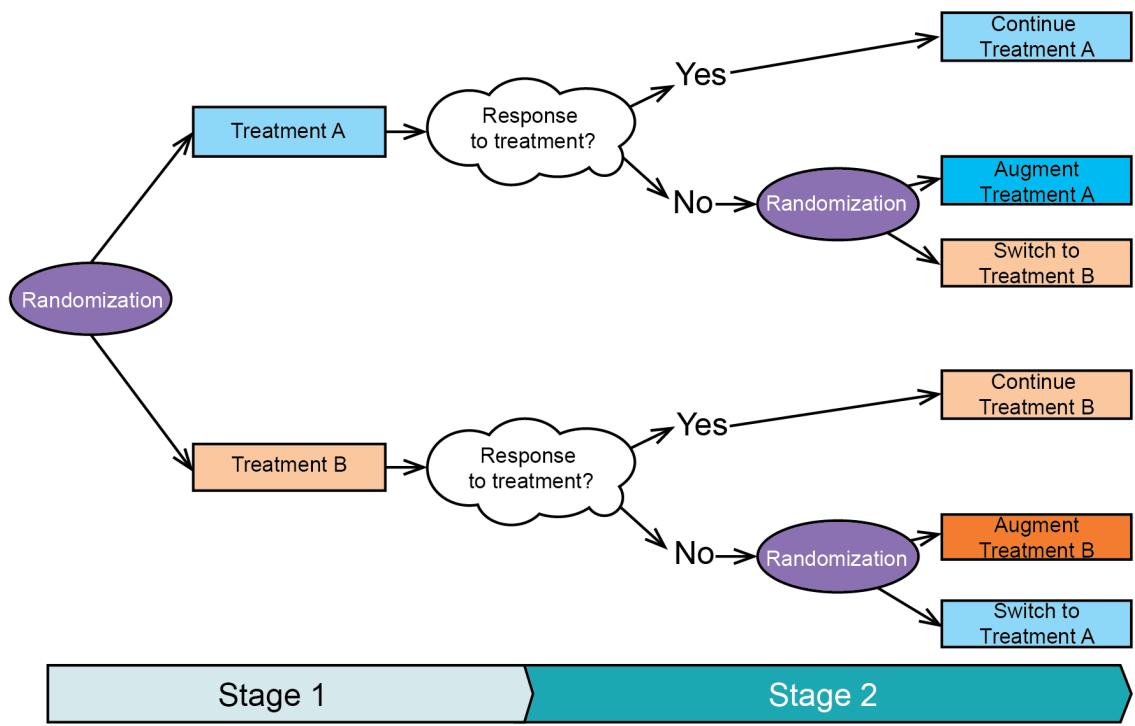

# Analysis plan for A Review of Methods and Practices for Missing Data in Sequential Multiple Assignment Randomized Trials (SMARTs): An Ancillary Study of a Scoping Review

**Manuscript authors:** Nikki L. B. Freeman[1,2,3], Chenyao Yu[7], Margaret Hoch, Sydney E. Browder[4], Bradley G. Hammill[2,5], Avi Kenny[1,6], Kevin J. Anstrom[8,9], and Michael R. Kosorok[3,8]

1. Department of Biostatistics and Bioinformatics, Duke University, Durham, NC
2. Duke Clinical Research Institute, Duke University, Durham, NC
3. Center for A.I. and Public Health, University of North Carolina at Chapel Hill, NC
4. Department of Epidemiology, University of North Carolina at Chapel Hill, Chapel Hill, NC
5. Department of Population Health Science, Duke University, Durham, NC
6. Duke Global Health Institute, Duke University, Durham, NC
7. Department of Statistical Science, Duke University, Durham, NC
8. Department of Biostatistics, University of North Carolina at Chapel Hill, Chapel Hill, NC
9. Collaborative Studies Coordinating Center, University of North Carolina at Chapel Hill, NC

**Analysis plan author:** Nikki L. B. Freeman (nikki.freeman@duke.edu)

**1. Background and Rationale**

Sequential multiple assignment randomized trials (SMARTs) are multi-stage clinical trial designs in which participants may be randomized to treatments at two or more sequential decision points, with later randomizations potentially depending on response to earlier treatments. SMARTs are uniquely designed to generate data for estimating dynamic treatment regimes (DTRs) and comparing embedded adaptive treatment strategies (Kidwell and Almirall 2023; Kosorok and Moodie 2015).

Missing data poses a particularly acute threat to SMART analyses because of the sequential treatment structure, response-dependent re-randomization, and multiple interrelated analytic targets. Data may be missing not only for the primary endpoint but also for intermediate responder/nonresponder classifications, second-stage treatment assignments, and time-varying covariates. Each type of missingness has distinct implications depending on the analytic target. Despite this, the methodological literature on handling missing data in SMARTs is sparse, and it is unknown how missing data is reported and handled in practice.

In a recently published scoping review (Freeman et al. 2025), we identified 89 SMART studies with published protocol and/or primary analysis papers. While that review focused broadly on design characteristics and reporting practices, it did not systematically examine missing data. The parent review's extraction form included two fields related to missing data (whether missingness was accounted for, and what methods were used), but these were not analyzed in depth. This ancillary study conducts a focused secondary analysis of missing data reporting and handling in the SMART primary analyses identified in the parent review.

**2. Objectives**

This study has two objectives:

1. Review the current state of statistical methods for handling missing data in SMARTs. (Narrative review; no preregistration of search strategy required.)
2. Characterize how missing data is reported and handled in published SMART results, drawing on the cohort of SMARTs identified in the parent scoping review. (Empirical secondary analysis; preregistered below.)

**3. Study Sample**

The study sample consists of the 30 SMART primary analysis papers identified in the parent scoping review: 16 primary analysis-only papers and 14 primary analysis papers with corresponding protocol papers. For the 14 studies with both a protocol and a primary analysis paper, we will also extract planned missing-data approaches from the protocol papers to enable a planned-versus-implemented concordance analysis.

The parent scoping review also examined protocol and design papers for SMARTs. Whether plans for missing data analysis were reported in the protocol-only papers (n = 59) was recorded during the parent scoping review extraction process. This information has not previously been reported on and will be included in this analysis. Methods for the data extraction process are available in the parent scoping review protocol and manuscript (Freeman et al. 2023, 2025).

No additional search will be conducted beyond the parent review. The parent review searched PubMed, Scopus, EMBASE, and CENTRAL through February 9, 2024. The full search strategy and inclusion/exclusion criteria are described in the parent study and its registered protocol on OSF (Freeman et al. 2023).

**4. Extraction Form**

We developed an expanded extraction form with 41 fields organized into seven domains. The complete extraction form and codebook are attached as Supplementary File 1. Below, we summarize the domains and their primary fields.

**Domain 1: Attrition and study flow (6 fields)**

Whether a CONSORT-style flow diagram was provided; number included in primary analysis; overall attrition rate; whether attrition was reported by treatment arm; whether stage-specific attrition was reported (i.e., how many participants reached the second-stage randomization); whether differential attrition was assessed.

**Domain 2: Types of missing data reported (6 fields)**

Which types of missing data were explicitly reported or discussed (missing primary outcome, missing secondary/interim outcome, missing responder/nonresponder status, missing second-stage treatment assignment, missing covariates, general "dropout" without specifics, or not discussed); rate of primary outcome missingness; whether responder status missingness was specifically reported; whether treatment assignment missingness was specifically reported; whether the missingness pattern (monotone vs. non-monotone) was discussed; timing of dropout.

**Domain 3: Missing data mechanism (2 fields)**

Whether the assumed missing data mechanism (MCAR/MAR/MNAR) was discussed; whether justification was provided for the assumed mechanism.

**Domain 4: Methods used to handle missing data (5 fields)**

Whether any method was used beyond complete case analysis; the primary method used (classified as: complete case/listwise deletion, last observation carried forward, multiple imputation, single imputation, inverse probability weighting, mixed models/likelihood-based, doubly robust/AIPW, SMART-specific method, other, or none described); additional methods used; whether a SMART-specific method was used (e.g., Shortreed et al. (2014), Dong et al. (2020)); details of the multiple imputation procedure if applicable (number of imputations, imputation model described, whether the SMART's sequential structure was accounted for).

**Domain 5: Sensitivity analyses (3 fields)**

Whether sensitivity analyses related to missing data were conducted; type of sensitivity analysis; whether sensitivity results were discussed in terms of robustness.

**Domain 6: Reporting and discussion (3 fields)**

Whether missing data was discussed as a study limitation; whether the potential impact of missingness on validity was discussed; whether missingness was discussed in the context of the SMART design specifically (e.g., impact on regime comparisons, DTR estimation, responder-dependent dropout).

**Domain 7: Protocol concordance (4 fields)**

Applicable to the 14 studies with both paper types. Whether the study has a protocol paper; planned missing data methods from the protocol; planned sensitivity analyses from the protocol; concordance between planned and implemented approaches (full, partial, or none).

Additionally, extractor metadata fields record extraction difficulty, extractor and verifier initials, and date.

**5. Extraction Procedure**

**5.1 Pre-population**

Study identification fields (columns A–G: SR number, first author/year, SMART name, disease area, pilot status, N randomized, number of stages) will be pre-populated from the parent scoping review's extraction database. For the 14 studies with protocol papers, planned missing data fields will be pre-populated from the parent extraction (columns AX–AZ).

**5.2 AI-assisted extraction**

We will use a large language model (Anthropic 2026) to assist with initial data extraction. The AI extraction process follows a structured prompt (attached in Supplementary File 1, Sheet 3) that specifies all extraction fields, their definitions, response options, and instructions for handling ambiguous cases. The AI returns structured output (JSON format) for each paper.

**5.3 Calibration**

Before full-scale extraction, we will calibrate the AI-assisted approach on 6 papers selected to represent varying levels of expected missing data reporting. For these papers, an author will independently perform full manual extraction. AI extractions will then be compared against manual extractions field by field for agreement between the human extractor and the AI extraction. Fields will be classified into reliability tiers (high if concordant among 5 or more papers, moderate if concordant on 4 or more papers, and low if concordant on 3 or less papers). This classification will determine the verification protocol for subsequent papers. If the AI extraction prompt requires refinement based on calibration, modifications will be documented in the decision log before proceeding.

**5.4 Tiered verification**

Once calibration is completed and any changes to the prompt are finalized, all papers included in the study will be extracted using AI-assistance. For the papers not included in the calibration step, AI-generated extractions will be verified by a human reviewer according to the reliability tiers established during calibration:

- High-reliability fields: Spot-checked in a random 20% sample of papers
- Moderate-reliability fields: Verified for every paper by targeted comparison with the source text
- Low-reliability fields: Manually extracted from the source paper, with AI output used only as a starting reference

For all fields where the AI reports information as "not reported" or "not discussed," the reviewer will confirm by checking the paper's methods section, results section, limitations section, and supplementary materials.

**5.5 Second-reviewer verification**

A second reviewer will independently verify a random sample of 6 papers against the source papers. Agreement will be based on key fields: primary missing data method (column W), whether a SMART-specific method was used (column Y), whether sensitivity analyses were conducted (column AA), and overall attrition rate (column J). If discordance exceeds 30% in these specific fields, we will specify a set of key fields for a second reviewer to manually extract from all papers. Discrepancies will be resolved by consensus.

**5.6 Protocol concordance**

For the 14 studies with both paper types, concordance between planned and implemented missing data approaches will be assessed by a human reviewer (not AI-assisted). Concordance will be classified as full, partial, or none, with a brief narrative description of any discrepancies.

**6. Planned Analyses**

**6.1 Primary descriptive analysis**

For each extraction field, when appropriate, we will compute descriptive statistics:

- Categorical fields: Frequencies and percentages with exact 95% confidence intervals where informative
- Continuous fields (e.g., attrition rates): Median and range

Results will be presented for the full sample of 30 primary analysis papers. Where sample sizes permit, we will also present results stratified by paper type (primary analysis-only, n = 16, vs. protocol+primary analysis, n = 14) and by pilot status.

**6.2 Planned summary tables**

We plan to construct the following tables for the manuscript:

- Table: Attrition and study flow reporting (Domain 1 results)
- Table: Methods used to handle missing data (Domain 4 results, with Domains 2–3 context)
- Table: Protocol concordance (Domain 7 results for the 14 applicable studies)
- Supplementary table: Study-level extraction results for all 30 papers across key fields

Additional tables may be constructed based on the findings.

**6.3 Protocol concordance analysis**

For the 14 studies with both paper types, we will present a descriptive comparison of planned versus implemented approaches to missing data. This will be reported as a study-level table showing, for each study, the planned method (from the protocol), the implemented method (from the primary analysis), and the concordance classification. We will summarize the overall concordance rate and describe common patterns of discrepancy.

**6.4 Extraction difficulty as a finding**

The distribution of extraction difficulty ratings (column AL: 1 = clearly reported, 2 = scattered, 3 = minimal/inferred, 4 = essentially not addressed) will be reported as a secondary finding about the state of missing data reporting transparency in SMART primary analyses.

**6.5 Analyses not planned**

Given the descriptive nature of this study and the small sample size (n = 30), we do not plan inferential statistical tests. We do not plan meta-analysis of attrition rates or effect sizes. We do not plan to contact study authors for additional information; this study assesses what was reported in published papers.

**7. Deviations from This Plan**

Any deviations from this preregistered plan will be documented and reported transparently in the manuscript. This includes changes to the extraction form after calibration (e.g., fields added, modified, or dropped), changes to the verification protocol, and any unplanned analyses. We anticipate that the calibration phase may lead to refinements of the AI extraction prompt; such refinements will be documented in the decision log and the final prompt will be made available with the supplementary materials.

**8. Reporting**

The parent scoping review adhered to the Preferred Reporting Items for Systematic Reviews and Meta-Analyses extension for Scoping Reviews (PRISMA-ScR) (Tricco et al. 2018). For the synthesis of this ancillary study, we will additionally consider the Synthesis Without Meta-analysis (SWiM) reporting guideline (Campbell et al. 2020). The primary manuscript may be posted on a preprint server prior to submission for publication.

**9. Supplementary Files**

The following supplementary files are attached to this preregistration:

- Supplementary File 1: Extraction form workbook (Excel), containing three sheets: (a) the extraction form with 41 fields and response option instructions, (b) a codebook with field definitions, response options, and granularity rationale, and (c) the AI extraction prompt template.
- The parent scoping review's protocol, search strategy, and extraction forms are available through the parent study's OSF registration.

**10. Timeline**

Note that other than the preregistration date, the timeline represents target dates. If challenges are encountered in the analysis process, the date on which these activities are actually executed will be later than posted here.

Preregistration: 29 March 2026

Calibration (Phase 1): 2 April 2026 – 4 April 2026

AI-assisted extraction and verification (Phase 2): 6 April 2026 – 10 April 2026

Quality control and second-reviewer verification (Phase 4): 11 April 2026 – 14 April 2026

Analysis and manuscript completion: 15 April 2026 – 18 April 2026

Preprint posting: 19 April 2026

**Funding and Conflicts of Interest**

The author of this analysis plan (NLBF) works on research projects funded by the National Institutes of Health, Novartis, Amgen, and the American Diabetes Association.

**Supplementary Table 1.** Included studies and study characteristics

| ID | First Author, Year | Title | Disease Area | Pilot? | N Randomized (Stage 1) | Number of Stages | Has Protocol Paper |
|---|---|---|---|---|---|---|---|
| 1099 | Butzer JF et al., 2023 | Randomized Trial of Two Exercise Programs to Increase Physical Activity and Health-Related Quality of Life for Persons With Spinal Cord Injury | Neurological disease or injury | No | 79 | 2 | No |
| 1282 | Czyz EK et al., 2021 | Adaptive intervention for prevention of adolescent suicidal behavior after hospitalization: a pilot sequential multiple assignment randomized trial | Mental and behavioral health | Yes | 80 | 2 | No |
| 1271 | Fatori D et al., 2018 | Adaptive treatment strategies for children and adolescents with Obsessive-Compulsive Disorder: A sequential multiple assignment randomized trial | Mental and behavioral health | No | 83 | 2 | No |
| 1232 | Fortney JC et al., 2021 | Comparison of Teleintegrated Care and Telereferral Care for Treating Complex Psychiatric Disorders in Primary Care: A Pragmatic Randomized Comparative Effectiveness Trial | Substance use | No | 1004 | 2 | Yes |
| 1081 | Gao K et al., 2020 | Sequential Multiple Assignment Randomized Treatment (SMART) for Bipolar Disorder at Any Phase of Illness and at least Mild Symptom Severity | Mental and behavioral health | No | 112 | 2 | No |
| 1273 | Geng EH et al., 2023 | Adaptive Strategies for Retention in Care among Persons Living with HIV | HIV | No | 1809 | 2 | No |
| 1014 | Gonze BB et al., 2020 | Use of a Smartphone App to Increase Physical Activity Levels in Insufficiently Active Adults: Feasibility Sequential Multiple Assignment Randomized Trial (SMART) | Mental and behavioral health | No | 18 | 2 | Yes |
| 1223 | Gunlicks-Stoessel M et al., 2019 | Critical Decision Points for Augmenting Interpersonal Psychotherapy for Depressed Adolescents: A Pilot Sequential Multiple Assignment Randomized Trial | Mental and behavioral health | Yes | 40 | 2 | Yes |
| 1008 | Igudesman D et al., 2023 | Weight management in young adults with type 1 diabetes: The advancing care for type 1 diabetes and obesity network sequential multiple assignment randomized trial pilot results | Overweight and obesity | Yes | 38 | 3 | Yes |
| 1177 | Karp JF et al., 2019 | Improving Patient Reported Outcomes and Preventing Depression and Anxiety in Older Adults With Knee Osteoarthritis: Results of a Sequenced Multiple Assignment Randomized Trial (SMART) Study | Osteoarthritis | Yes | 99 | 2 | Yes |
| 1124 | Kruse GR et al., 2023 | A pilot adaptive trial of text messages, mailed nicotine replacement therapy, and telephone coaching among primary care patients who smoke | Substance use | Yes | 35 | 2 | No |
| 1269 | Lambert SD et al., 2022 | Adaptive web-based stress management programs among adults with a cardiovascular disease: A pilot Sequential Multiple Assignment Randomized Trial (SMART) | Mental and behavioral health | Yes | 59 | 2 | No |
| 1203 | McKay JR et al., 2015 | Effect of patient choice in an adaptive sequential randomization trial of treatment for alcohol and cocaine dependence | Substance use | No | 500 | 2 | No |
| 1200 | Morgenstern J et al., 2021 | An efficacy trial of adaptive interventions for alcohol use disorder | Substance use | No | 133 | 2 | No |
| 1216 | Morin CM et al., 2020 | Effectiveness of Sequential Psychological and Medication Therapies for Insomnia Disorder: A Randomized Clinical Trial | Sleep | No | 211 | 2 | Yes |
| 1298 | Mustanski B et al., 2023 | Effectiveness of the SMART Sex Ed program among 13-18 year old English and Spanish speaking adolescent men who have sex with men | HIV | No | 983 | 2 | Yes |
| 1078 | Naar-King S et al., 2016 | Sequential Multiple Assignment Randomized Trial (SMART) to Construct Weight Loss Interventions for African American Adolescents | Overweight and obesity | No | 18 | 2 | No |
| 1154 | Patrick ME et al., 2021 | Main outcomes of M-bridge: A sequential multiple assignment randomized trial (SMART) for developing an adaptive preventive intervention for college drinking | Substance use | No | 891 | 2 | Yes |
| 1020 | Pelham Jr WE et al., 2016 | Treatment Sequencing for Childhood ADHD: A Multiple-Randomization Study of Adaptive Medication and Behavioral Interventions | Mental and behavioral health | No | 152 | 2 | No |
| 1219 | Pistorello J et al., 2017 | Developing Adaptive Treatment Strategies to Address Suicidal Risk in College Students: A Pilot Sequential, Multiple Assignment, Randomized Trial (SMART) | Mental and behavioral health | Yes | 62 | 2 | No |
| 1062 | Sauer-Zavala S et al., 2022 | A SMART approach to personalized care: preliminary data on how to select and sequence skills in transdiagnostic CBT | Mental and behavioral health | Yes | 70 | 2 | No |
| 1005 | Schlam TR et al., 2024 | What to do after smoking relapse? A sequential multiple assignment randomized trial of chronic care smoking treatments | Substance use | No | 1154 | 2 | No |
| 1228 | Schmitz JM et al., 2024 | Contingency management plus acceptance and commitment therapy for initial cocaine abstinence: Results of a sequential multiple assignment randomized trial (SMART) | Substance use | No | 118 | 2 | Yes |
| 1121 | Schoenfelder EN et al., 2019 | Piloting a Sequential, Multiple Assignment, Randomized Trial for Mothers with Attention-Deficit/Hyperactivity Disorder and Their At-Risk Young Children | Mental and behavioral health | Yes | 35 | 2 | Yes |
| 1246 | Sherwood NE et al., 2022 | BestFIT Sequential Multiple Assignment Randomized Trial Results: A SMART Approach to Developing Individualized Weight Loss Treatment Sequences | Overweight and obesity | No | 468 | 2 | Yes |
| 1075 | Sikorskii A et al., 2023 | A Sequential Multiple Assignment Randomized Trial of Symptom Management After Chemotherapy | Cancer | No | 451 | 2 | No |
| 1112 | Smith SN et al., 2022 | Primary aim results of a clustered SMART for developing a school-level, adaptive implementation strategy to support CBT delivery at high schools in Michigan | Mental and behavioral health | No | 94 schools (169 SPs) | 2 | Yes |
| 1250 | Somers TJ et al., 2023 | Behavioral cancer pain intervention dosing: results of a Sequential Multiple Assignment Randomized Trial | Pain | No | 327 | 2 | Yes |
| 1002 | Stanger C et al., 2020 | Working memory training and high magnitude incentives for youth cannabis use: A SMART pilot trial | Substance use | Yes | 59 | 2 | No |
| 1097 | Wyatt G et al., 2021 | Reflexology and meditative practices for symptom management among people with cancer: Results from a sequential multiple assignment randomized trial | Cancer | No | 347 | 2 | Yes |

**Supplementary Table 2.** Full extraction for attrition and study flow

| ID | First Author, Year | Title | Pilot? | CONSORT Diagram | Overall Attrition Rate | Attrition between stage 1 and stage 2 |
|---|---|---|---|---|---|---|
| 1112 | Smith SN et al., 2022 | Primary aim results of a clustered SMART for developing a school-level, adaptive implementation strategy to support CBT delivery at high schools in Michigan | No | Yes | 0.6 | 0 |
| 1282 | Czyz EK et al., 2021 | Adaptive intervention for prevention of adolescent suicidal behavior after hospitalization: a pilot sequential multiple assignment randomized trial | Yes | Yes | 1.25 | 0 |
| 1232 | Fortney JC et al., 2021 | Comparison of Teleintegrated Care and Telereferral Care for Treating Complex Psychiatric Disorders in Primary Care: A Pragmatic Randomized Comparative Effectiveness Trial | No | Yes | 36.5 | 0 |
| 1020 | Pelham Jr WE et al., 2016 | Treatment Sequencing for Childhood ADHD: A Multiple-Randomization Study of Adaptive Medication and Behavioral Interventions | No | Yes | 3.9 | Not reported |
| 1250 | Somers TJ et al., 2023 | Behavioral cancer pain intervention dosing: results of a Sequential Multiple Assignment Randomized Trial | No | Yes | 8.9 | 4.3 |
| 1246 | Sherwood NE et al., 2022 | BestFIT Sequential Multiple Assignment Randomized Trial Results: A SMART Approach to Developing Individualized Weight Loss Treatment Sequences | No | Yes | Indeterminant | 5.6 |
| 1228 | Schmitz JM et al., 2024 | Contingency management plus acceptance and commitment therapy for initial cocaine abstinence: Results of a sequential multiple assignment randomized trial (SMART) | No | Yes | 24.5 | 5.9 |
| 1216 | Morin CM et al., 2020 | Effectiveness of Sequential Psychological and Medication Therapies for Insomnia Disorder: A Randomized Clinical Trial | No | Yes | Indeterminant | Indeterminant |
| 1223 | Gunlicks-Stoessel M et al., 2019 | Critical Decision Points for Augmenting Interpersonal Psychotherapy for Depressed Adolescents: A Pilot Sequential Multiple Assignment Randomized Trial | Yes | Yes | 17.5 | 2.5 |
| 1075 | Sikorskii A et al., 2023 | A Sequential Multiple Assignment Randomized Trial of Symptom Management After Chemotherapy | No | Yes | 15.2 | 6 |
| 1121 | Schoenfelder EN et al., 2019 | Piloting a Sequential, Multiple Assignment, Randomized Trial for Mothers with Attention-Deficit/Hyperactivity Disorder and Their At-Risk Young Children | Yes | Yes | 11.4 | 2.9 |
| 1273 | Geng EH et al., 2023 | Adaptive Strategies for Retention in Care among Persons Living with HIV | No | Yes | 9.4 | 7.7 |
| 1078 | Naar-King S et al., 2016 | Sequential Multiple Assignment Randomized Trial (SMART) to Construct Weight Loss Interventions for African American Adolescents | No | Yes | 11.8 | 8.6 |
| 1177 | Karp JF et al., 2019 | Improving Patient Reported Outcomes and Preventing Depression and Anxiety in Older Adults With Knee | Yes | Yes | 20.2 | 10.1 |
| | | Osteoarthritis: Results of a Sequenced Multiple Assignment Randomized Trial (SMART) Study | | | | |
| 1200 | Morgenstern J et al., 2021 | An efficacy trial of adaptive interventions for alcohol use disorder | No | Yes | 10.5 | 9.8 |
| 1124 | Kruse GR et al., 2023 | A pilot adaptive trial of text messages, mailed nicotine replacement therapy, and telephone coaching among primary care patients who smoke | Yes | Yes | 14.3 | 11.4 |
| 1269 | Lambert SD et al., 2022 | Adaptive web-based stress management programs among adults with a cardiovascular disease: A pilot Sequential Multiple Assignment Randomized Trial (SMART) | Yes | Yes | 32.2 | 15.3 |
| 1062 | Sauer-Zavala S et al., 2022 | A SMART approach to personalized care: preliminary data on how to select and sequence skills in transdiagnostic CBT | Yes | Yes | Indeterminant | Not reported |
| 1271 | Fatori D et al., 2018 | Adaptive treatment strategies for children and adolescents with Obsessive-Compulsive Disorder: A sequential multiple assignment randomized trial | No | Yes | 24.1 | 13.2 |
| 1154 | Patrick ME et al., 2021 | Main outcomes of M-bridge: A sequential multiple assignment randomized trial (SMART) for developing an adaptive preventive intervention for college drinking | No | Yes | 17.5 | 13.5 |
| 1298 | Mustanski B et al., 2023 | Effectiveness of the SMART Sex Ed program among 13-18 year old English and Spanish speaking adolescent men who have sex with men | No | No | 24.7 | Not applicable |
| 1203 | McKay JR et al., 2015 | Effect of patient choice in an adaptive sequential randomization trial of treatment for alcohol and cocaine dependence | No | Yes | Indeterminant | 15.9 |
| 1008 | Igudesman D et al., 2023 | Weight management in young adults with type 1 diabetes: The advancing care for type 1 diabetes and obesity network sequential multiple assignment randomized trial pilot results | Yes | No (only reported first randomization) | 25.5 | Not applicable |
| 1002 | Stanger C et al., 2020 | Working memory training and high magnitude incentives for youth cannabis use: A SMART pilot trial | Yes | Yes | 18.6 | 18.6 |
| 1099 | Butzer JF et al., 2023 | Randomized Trial of Two Exercise Programs to Increase Physical Activity and Health-Related Quality of Life for Persons With Spinal Cord Injury | No | Yes | Indeterminant | 16.5 |
| 1219 | Pistorello J et al., 2017 | Developing Adaptive Treatment Strategies to Address Suicidal Risk in College Students: A Pilot Sequential, Multiple Assignment, Randomized Trial (SMART) | Yes | Yes | 27.4 | 22.6 |
| 1097 | Wyatt G et al., 2021 | Reflexology and meditative practices for symptom management among people with cancer: Results from a sequential multiple assignment randomized trial | No | Yes | Indeterminant | 20.5 |
| 1005 | Schlam TR et al., 2024 | What to do after smoking relapse? A sequential multiple assignment randomized trial of chronic care smoking treatments | No | Yes | 43.8 | 24.6 |
| 1014 | Gonze BB et al., 2020 | Use of a Smartphone App to Increase Physical Activity Levels in Insufficiently Active Adults: Feasibility | No | Yes | 33.3 | 27.8 |

| | | Sequential Multiple Assignment Randomized Trial (SMART) | | | | |
|---|---|---|---|---|---|---|
| **1081** | Gao K et al., 2020 | Sequential Multiple Assignment Randomized Treatment (SMART) for Bipolar Disorder at Any Phase of Illness and at least Mild Symptom Severity | No | Yes | 56.5 | 34.8 |

**Supplementary Table 3**. Full extraction for missing data mechanism

| ID | First Author, Year | Title | Mechanism Discussed | Mechanism Justification |
|---|---|---|---|---|
| 1008 | Igudesman D et al., 2023 | Weight management in young adults with type 1 diabetes: The advancing care for type 1 diabetes and obesity network sequential multiple assignment randomized trial pilot results | Not discussed | NA |
| 1014 | Gonze BB et al., 2020 | Use of a Smartphone App to Increase Physical Activity Levels in Insufficiently Active Adults: Feasibility Sequential Multiple Assignment Randomized Trial (SMART) | Not discussed | NA |
| 1097 | Wyatt G et al., 2021 | Reflexology and meditative practices for symptom management among people with cancer: Results from a sequential multiple assignment randomized trial | MAR | Baseline values of the outcomes of drop-outs were compared by randomized condition to evaluate the assumption |
| 1112 | Smith SN et al., 2022 | Primary aim results of a clustered SMART for developing a school-level, adaptive implementation strategy to support CBT delivery at high schools in Michigan | Not discussed | NA |
| 1121 | Schoenfelder EN et al., 2019 | Piloting a Sequential, Multiple Assignment, Randomized Trial for Mothers with Attention-Deficit/Hyperactivity Disorder and Their At-Risk Young Children | Not discussed | NA |
| 1154 | Patrick ME et al., 2021 | Main outcomes of M-bridge: A sequential multiple assignment randomized trial (SMART) for developing an adaptive preventive intervention for college drinking | MAR - Implicit based on methods used | No |
| 1177 | Karp JF et al., 2019 | Improving Patient Reported Outcomes and Preventing Depression and Anxiety in Older Adults With Knee Osteoarthritis: Results of a Sequenced Multiple Assignment Randomized Trial (SMART) Study | MAR | Authors justified MAR assumption based on: 1) minimal missing data, and 2) similar rates across randomized groups. No formal testing or sensitivity analysis to support this assumption |
| 1216 | Morin CM et al., 2020 | Effectiveness of Sequential Psychological and Medication Therapies for Insomnia Disorder: A Randomized Clinical Trial | MAR - Implicit based on methods used | NA |
| 1223 | Gunlicks-Stoessel M et al., 2019 | Critical Decision Points for Augmenting Interpersonal Psychotherapy for Depressed Adolescents: A Pilot Sequential Multiple Assignment Randomized Trial | Not discussed | No explicit justification, though differential attrition tests provide indirect support |
| 1228 | Schmitz JM et al., 2024 | Contingency management plus acceptance and commitment therapy for initial cocaine abstinence: Results of a sequential multiple assignment randomized trial (SMART) | MNAR - Implicitly addressed through conservative assumptions regarding missingness | NA |
| 1232 | Fortney JC et al., 2021 | Comparison of Teleintegrated Care and Telereferral Care for Treating Complex Psychiatric Disorders in Primary Care: A Pragmatic Randomized Comparative Effectiveness Trial | MAR | Authors examined correlations between loss to follow-up and baseline characteristics and conducted sensitivity analyses for violations of MAR assumption. |
| 1246 | Sherwood NE et al., 2022 | BestFIT Sequential Multiple Assignment Randomized Trial Results: A SMART Approach to Developing Individualized Weight Loss Treatment Sequences | MAR | Partial - stated that imputation model assumed MAR conditional on covariates and auxiliary variables (education, income, coach, intervention completion), but no additional justification provided |
| 1250 | Somers TJ et al., 2023 | Behavioral cancer pain intervention dosing: results of a Sequential Multiple Assignment Randomized Trial | Not discussed | NA |
| 1298 | Mustanski B et al., 2023 | Effectiveness of the SMART Sex Ed program among 13-18 year old English and Spanish speaking adolescent men who have sex with men | MAR - Implicit based on methods used | Authors state they 'did not observe systematic differences between participants who did and did not complete three-month follow-up with regard to demographic factors', suggesting missingness was unrelated to observed demographics, but no formal testing provided |
| 1002 | Stanger C et al., 2020 | Working memory training and high magnitude incentives for youth cannabis use: A SMART pilot trial | MNAR - Implicitly addressed through conservative assumptions regarding missingness | NA |
| 1005 | Schlam TR et al., 2024 | What to do after smoking relapse? A sequential multiple assignment randomized trial of chronic care smoking treatments | MNAR - Implicitly addressed through conservative assumptions regarding missingness | NA |
| 1020 | Pelham Jr WE et al., 2016 | Treatment Sequencing for Childhood ADHD: A Multiple-Randomization Study of Adaptive Medication and Behavioral Interventions | MAR | Authors state MAR is plausible 'given the inclusion of a large number of covariates, including baseline measures of outcome variables; measures' values at earlier waves are typically the best predictors of missing values at later waves in a longitudinal design.' |
| 1062 | Sauer-Zavala S et al., 2022 | A SMART approach to personalized care: preliminary data on how to select and sequence skills in transdiagnostic CBT | Not discussed | NA |
| 1075 | Sikorskii A et al., 2023 | A Sequential Multiple Assignment Randomized Trial of Symptom Management After Chemotherapy | Not discussed | NA |
| 1078 | Naar-King S et al., 2016 | Sequential Multiple Assignment Randomized Trial (SMART) to Construct Weight Loss Interventions for African American Adolescents | MAR | No |
| 1081 | Gao K et al., 2020 | Sequential Multiple Assignment Randomized Treatment (SMART) for Bipolar Disorder at Any Phase of Illness and at least Mild Symptom Severity | Not discussed | NA |
| 1099 | Butzer JF et al., 2023 | Randomized Trial of Two Exercise Programs to Increase Physical Activity and Health- | Not discussed | NA |
| | | Related Quality of Life for Persons With Spinal Cord Injury | | |
| **1124** | Kruse GR et al., 2023 | A pilot adaptive trial of text messages, mailed nicotine replacement therapy, and telephone coaching among primary care patients who smoke | Not discussed | NA |
| **1200** | Morgenstern J et al., 2021 | An efficacy trial of adaptive interventions for alcohol use disorder | MAR - Implicit based on methods used | NA |
| **1203** | McKay JR et al., 2015 | Effect of patient choice in an adaptive sequential randomization trial of treatment for alcohol and cocaine dependence | MAR - Implicit based on methods used | NA |
| **1219** | Pistorello J et al., 2017 | Developing Adaptive Treatment Strategies to Address Suicidal Risk in College Students: A Pilot Sequential, Multiple Assignment, Randomized Trial (SMART) | Not discussed | NA |
| **1269** | Lambert SD et al., 2022 | Adaptive web-based stress management programs among adults with a cardiovascular disease: A pilot Sequential Multiple Assignment Randomized Trial (SMART) | Not discussed | NA |
| **1271** | Fatori D et al., 2018 | Adaptive treatment strategies for children and adolescents with Obsessive-Compulsive Disorder: A sequential multiple assignment randomized trial | Not discussed | NA |
| **1273** | Geng EH et al., 2023 | Adaptive Strategies for Retention in Care among Persons Living with HIV | MAR - Implicit based on methods used | NA |
| **1282** | Czyz EK et al., 2021 | Adaptive intervention for prevention of adolescent suicidal behavior after hospitalization: a pilot sequential multiple assignment randomized trial | MAR - Implicit based on methods used | NA |

Supplementary Table 4. Full extraction for the missing data methods reported

| ID | First Author, Year | Title | Pilot? | Any Method Used | Primary Missing Data Method | Additional Methods | SMART-Specific Method Used |
|---|---|---|---|---|---|---|---|
| 1008 | Igudesman D et al., 2023 | Weight management in young adults with type 1 diabetes: The advancing care for type 1 diabetes and obesity network sequential multiple assignment randomized trial pilot results | Yes | Yes | Complete case | Single imputation | No |
| 1014 | Gonze BB et al., 2020 | Use of a Smartphone App to Increase Physical Activity Levels in Insufficiently Active Adults: Feasibility Sequential Multiple Assignment Randomized Trial (SMART) | No | Yes | Complete case | None | No |
| 1097 | Wyatt G et al., 2021 | Reflexology and meditative practices for symptom management among people with cancer: Results from a sequential multiple assignment randomized trial | No | Yes | Mixed models/likelihood-based | Supplementary analysis of dropout baseline characteristics | No |
| 1112 | Smith SN et al., 2022 | Primary aim results of a clustered SMART for developing a school-level, adaptive implementation strategy to support CBT delivery at high schools in Michigan | No | Yes | Multiple imputation | Analyses also run without multiply-imputed data as sensitivity/comparison | Yes — Shortreed et al. 2014 multiple imputation strategy for SMARTs is cited as the basis for the MI approach |
| 1121 | Schoenfelder EN et al., 2019 | Piloting a Sequential, Multiple Assignment, Randomized Trial for Mothers with Attention-Deficit/Hyperactivity Disorder and Their At-Risk Young Children | Yes | No | None described | None | No |
| 1154 | Patrick ME et al., 2021 | Main outcomes of M-bridge: A sequential multiple assignment randomized trial (SMART) for developing an adaptive preventive intervention for college drinking | No | Yes | Multiple imputation | None | No |
| 1177 | Karp JF et al., 2019 | Improving Patient Reported Outcomes and Preventing Depression and Anxiety in Older Adults With Knee Osteoarthritis: Results of a Sequenced Multiple Assignment Randomized Trial (SMART) Study | Yes | Yes | Mixed models/likelihood-based | None | No |
| 1216 | Morin CM et al., 2020 | Effectiveness of Sequential Psychological and Medication Therapies for Insomnia Disorder: A Randomized Clinical Trial | No | Yes | Inverse probability weighting | None | No |
| 1223 | Gunlicks-Stoessel M et al., 2019 | Critical Decision Points for Augmenting Interpersonal Psychotherapy for Depressed Adolescents: A Pilot Sequential Multiple Assignment Randomized Trial | Yes | Yes | Multiple imputation | None | No |
| 1228 | Schmitz JM et al., 2024 | Contingency management plus acceptance and commitment therapy for initial cocaine abstinence: Results of a sequential multiple assignment randomized trial (SMART) | No | Yes | Single imputation | Follow-up sensitivity analyses operationalized proportion cocaine- | No |
| | | | | | | negative UDS using a constant denominator (max possible tests) vs. denominator up to dropout | |
| 1232 | Fortney JC et al., 2021 | Comparison of Teleintegrated Care and Telereferral Care for Treating Complex Psychiatric Disorders in Primary Care: A Pragmatic Randomized Comparative Effectiveness Trial | No | Yes | Multiple imputation | Sensitivity analysis for MNAR violations calculating how primary outcome estimate changed under different proportions and effect sizes of nonignorable missingness | No |
| 1246 | Sherwood NE et al., 2022 | BestFIT Sequential Multiple Assignment Randomized Trial Results: A SMART Approach to Developing Individualized Weight Loss Treatment Sequences | No | Yes | Mixed models/likelihood-based | Yes - Multiple imputation used as sensitivity analysis. MCMC for sporadically missing 6-month values; fully conditional specification (regression method) for monotonically missing 6- and 18-month values. GEE with imputed data used for exploratory adaptive intervention analyses | No |
| 1250 | Somers TJ et al., 2023 | Behavioral cancer pain intervention dosing: results of a Sequential Multiple Assignment Randomized Trial | No | No | None described | None | No |
| 1298 | Mustanski B et al., 2023 | Effectiveness of the SMART Sex Ed program among 13-18 year old English and Spanish speaking adolescent men who have sex with men | No | Yes | Multiple imputation | None | No |
| 1002 | Stanger C et al., 2020 | Working memory training and high magnitude incentives for youth cannabis use: A SMART pilot trial | Yes | Yes | Single imputation | Sensitivity analysis for WCA using number of negative samples, treating missing specimens as missing rather than positive; same pattern of results observed. Mixed models used for VSWM analysis | No |
| | | | | | | (likelihood-based, implicitly handles MAR) | |
| **1005** | Schlam TR et al., 2024 | What to do after smoking relapse? A sequential multiple assignment randomized trial of chronic care smoking treatments | No | Yes | Single imputation | Yes - multiple imputation was conducted as a sensitivity analysis for the primary outcome. | No |
| **1020** | Pelham Jr WE et al., 2016 | Treatment Sequencing for Childhood ADHD: A Multiple-Randomization Study of Adaptive Medication and Behavioral Interventions | No | Yes | Multiple imputation | Complete case for the 6 early withdrawals (excluded from analyses); multiple imputation applied only to the 146 completers for item/scale-level missingness on outcomes | No |
| **1062** | Sauer-Zavala S et al., 2022 | A SMART approach to personalized care: preliminary data on how to select and sequence skills in transdiagnostic CBT | Yes | Yes | Mixed models/likelihood-based | None | No |
| **1075** | Sikorskii A et al., 2023 | A Sequential Multiple Assignment Randomized Trial of Symptom Management After Chemotherapy | No | Yes | Mixed models/likelihood-based | Covariance adjustment for baseline values and variables identified in attrition analysis (variables with P<0.10 in attrition analysis were used as adjustment factors). Authors describe this as 'an additional safeguard for biases.' | No |
| **1078** | Naar-King S et al., 2016 | Sequential Multiple Assignment Randomized Trial (SMART) to Construct Weight Loss Interventions for African American Adolescents | No | Yes | Mixed models/likelihood-based | None | No |
| **1081** | Gao K et al., 2020 | Sequential Multiple Assignment Randomized Treatment (SMART) for Bipolar Disorder at Any Phase of Illness and at least Mild Symptom Severity | No | No | None described | None | No |
| **1099** | Butzer JF et al., 2023 | Randomized Trial of Two Exercise Programs to Increase Physical Activity and Health-Related Quality of Life for Persons With Spinal Cord Injury | No | No | None described | None | No |
| **1124** | Kruse GR et al., 2023 | A pilot adaptive trial of text messages, mailed nicotine replacement therapy, and telephone | Yes | No | None described | None | No |
| | | coaching among primary care patients who smoke | | | | | |
| 1200 | Morgenstern J et al., 2021 | An efficacy trial of adaptive interventions for alcohol use disorder | No | Yes | Mixed models/likelihood-based | None | No |
| 1203 | McKay JR et al., 2015 | Effect of patient choice in an adaptive sequential randomization trial of treatment for alcohol and cocaine dependence | No | Yes | Mixed models/likelihood-based | Yes, sensitivity analyses comparing GEE to mixed effects models, pattern mixture models, examination of baseline predictors of missingness | No |
| 1219 | Pistorello J et al., 2017 | Developing Adaptive Treatment Strategies to Address Suicidal Risk in College Students: A Pilot Sequential, Multiple Assignment, Randomized Trial (SMART) | Yes | No | None described | None described. Analyses appear to use available-case data (e.g., CSQ analyses report n=51 for S1 satisfaction) | No |
| 1269 | Lambert SD et al., 2022 | Adaptive web-based stress management programs among adults with a cardiovascular disease: A pilot Sequential Multiple Assignment Randomized Trial (SMART) | Yes | No | None described | None | No |
| 1271 | Fatori D et al., 2018 | Adaptive treatment strategies for children and adolescents with Obsessive-Compulsive Disorder: A sequential multiple assignment randomized trial | No | Yes | Single imputation | None | No |
| 1273 | Geng EH et al., 2023 | Adaptive Strategies for Retention in Care among Persons Living with HIV | No | Yes | Doubly robust (TMLE) | Adjudication process for classifying viral suppression when HIV RNA was missing. (Adjudication process for missing viral load measurements - 'persons without an HIV RNA measurement were classified as either treatment failure or missing using a standardized multidisciplinary adjudication process blinded to treatment group.' Kaplan-Meier estimator for Stage 1 retention analysis) | No |

| 1282 | Czyz EK et al., 2021 | Adaptive intervention for prevention of adolescent suicidal behavior after hospitalization: a pilot sequential multiple assignment randomized trial | Yes | Yes | Mixed models/likelihood-based | None | No |

Supplementary Table 5. Full extraction for protocol and primary missing data analysis concordance

| ID | First Author, Year | Title | Pilot? | Protocol Planned Methods | Protocol Planned Sensitivity | Concordance |
|---|---|---|---|---|---|---|
| 1014 | Gonze BB et al., 2020 | Use of a Smartphone App to Increase Physical Activity Levels in Insufficiently Active Adults: Feasibility Sequential Multiple Assignment Randomized Trial (SMART) | No | None | No | NA |
| 1097 | Wyatt G et al., 2021 | Reflexology and meditative practices for symptom management among people with cancer: Results from a sequential multiple assignment randomized trial | No | Mixed-models, likelihood based | Yes (used to investigate the robustness to MNAR) | Partially; sensitivity analyses were not used. |
| 1112 | Smith SN et al., 2022 | Primary aim results of a clustered SMART for developing a school-level, adaptive implementation strategy to support CBT delivery at high schools in Michigan | No | Multiple imputation | No | Yes - MI was implemented as planned; the primary paper additionally reported analyses with and without MI |
| 1154 | Patrick ME et al., 2021 | Main outcomes of M-bridge: A sequential multiple assignment randomized trial (SMART) for developing an adaptive preventive intervention for college drinking | No | None | No | NA |
| 1216 | Morin CM et al., 2020 | Effectiveness of Sequential Psychological and Medication Therapies for Insomnia Disorder: A Randomized Clinical Trial | No | Will investigate if data are MCAR, MAR, or MNAR; primary analysis method is robust to MCAR and MAR; No imputation will be performed and all observations will be included in analysis | Yes if there is MNAR | Somewhat; used likelihood-based approaches (MAR implicit) no assessment of MAR/MNAR or sensitivity analyses |
| 1228 | Schmitz JM et al., 2024 | Contingency management plus acceptance and commitment therapy for initial cocaine abstinence: Results of a sequential multiple assignment randomized trial (SMART) | No | conservative single imputation procedure for handling intermittent missing data | No | Conservative approach (missing == failure), sensitivity analysis of MI for missing baseline covariates; "missingness on this primary outcome was not associated with treatment condition at any phase", i.e., evaluated association between missingness and outcome |
| 1232 | Fortney JC et al., 2021 | Comparison of Teleintegrated Care and Telereferral Care for Treating Complex Psychiatric Disorders in Primary Care: A Pragmatic Randomized Comparative Effectiveness Trial | No | Multiple imputation | No | Partially: sensitivity analyses were used |
| 1246 | Sherwood NE et al., 2022 | BestFIT Sequential Multiple Assignment Randomized Trial Results: A SMART Approach to Developing Individualized Weight Loss Treatment Sequences | No | Mixed linear models for the primary and secondary analyses | No | Partially: the primary paper used the planned mixed-model framework, but added multiple imputation sensitivity analyses |
| 1250 | Somers TJ et al., 2023 | Behavioral cancer pain intervention dosing: results of a Sequential Multiple Assignment Randomized Trial | No | None | No | NA |

| 1298 | Mustanski B et al., 2023 | Effectiveness of the SMART Sex Ed program among 13-18 year old English and Spanish speaking adolescent men who have sex with men | No | None | No | No. No missing data approach discussed in the protocol, but the primary paper used MICE. |